\documentclass[12pt,a4paper,dvips]{article}
\usepackage{a4p}
\usepackage{cite,mcite}
\usepackage{graphicx}
\usepackage{physics }
\usepackage{l3_title,ifthen}
\usepackage{rotating}
\journalname{Phys. Lett. B}

\usepackage{Lep}
\Lep{2}

\preprint{99-117}
\date{August 09, 1999}

\newlength{\capindent}
\newlength{\capwidth}
\newlength{\figwidth}
\newcommand{\icaption}[2][!*!,!]{\hspace*{\capindent}%
  \begin{minipage}{\capwidth}
    \ifthenelse{\equal{#1}{!*!,!}}%
      {\caption{#2}}%
      {\caption[#1]{#2}}
  \end{minipage}}
\def\TeV{\ifmmode {\mathrm{\ Te\kern -0.1em V}}\else
                   \textrm{Te\kern -0.1em V}\fi}%
\journalname{Phys. Lett.}
\begin{document}
\setlength{\unitlength}{1mm}
\begin{titlepage}
\title{Search for Low Scale Gravity Effects
 in \boldmath{\epem} Collisions at LEP}
\author{The L3 Collaboration}
\begin{abstract}
Recent theories  propose that quantum gravity effects may be observable
at LEP energies via gravitons that couple to Standard
Model particles and propagate into extra spatial dimensions. 
The associated production of a graviton and a photon is
searched for as well as the effects of virtual graviton exchange in the
processes: $\epem\ra\gamma\gamma$, $\Zo\Zo$, $\Wp\Wm$,   $\mu^+\mu^-$, 
$\tau^+\tau^-$,  $\qqbar$ and $\epem$.
No evidence for this new interaction is found in the data sample
collected by the L3 detector at LEP at centre--of--mass energies up to
$183\GeV$. 
Limits close to $1\TeV$ on the scale of this new scenario of quantum gravity
are set.
\end{abstract}
\submitted
\end{titlepage}
%
%

\section{Introduction}
Two of the fundamental interactions of nature, the gravitational and
the electroweak, 
have widely differing characteristic scales, namely the Planck
($M_{Pl}\sim 10^{19}\GeV$) and the electroweak
($M_{ew}\sim 10^{2}\GeV$). 
The
Standard Model~\cite{sm_glashow} (SM)
successfully describes the electroweak interactions but leaves
unexplained the difference between these two scales.
While  electroweak interactions are probed at distances  of the order
of $M_{ew}^{-1}$, the gravitational force is studied only down
to distances of the order of a centimetre~\cite{expgravity}, thirty
three orders of magnitude  above its characteristic distance $M_{Pl}^{-1}$.

A recent theoretical scenario~\cite{arkani} proposes a modification
of the present 
understanding of the gravitational force interpreting a single scale,
$M_{ew}$, as the only fundamental one of nature. The 
known and tested behaviour of the gravitational force is accomodated by
the existence of $\delta$ new space dimensions of size $R$ such that
for the new scale of gravity, $M_D$, the following relation holds:
\begin{equation}
M_{Pl}^2 \sim R^\delta M_D^{\delta+2}.
\end{equation}

A single extra dimension, $\delta=1$, with $M_D \sim M_{ew}$ 
implies values of $R$ comparable to the dimensions of the solar system,
which is not allowed by the experimental knowledge of the gravitational
force, while starting from $\delta=2$ the corresponding dimensions of
$0.1 - 1$\,mm and below have yet to be
investigated\footnote{Severe limits on the
  $\delta=2$ scenario are  derived from SN1987A~\cite{arkani2},
  nonetheless a direct and complementary collider limit is desirable.}.

A consequence of this picture are spin 2 gravitons, $G$, that propagate in
$4+\delta$ dimensions, interacting with SM  particles in the
ordinary 4 dimensions with a sizeable strength, related to  
$M_D^{-1}$. This new interaction  is also
referred to as Low Scale Gravity (LSG).

The associated production  of a real
graviton and a  photon is searched for as well as 
the effects of virtual graviton exchange  in gauge boson or fermion
pair production.
Data collected in 1997 by the  L3
detector~\cite{l3_00}
in $\epem$ collisions at an average centre--of--mass energy $\sqrt{s}=182.7\GeV$
 denoted hereafter as $183\GeV$ are analysed. Data
samples  at $\sqrt{s}$ between $130\GeV$ and $172\GeV$ are also
considered for the $\gamma\gamma$ and $\rm{q\bar{q}}$ final states. 
Some of these channels are also investigated in Reference~\cite{opal1}.

%
%
\section{Real graviton production}

Real gravitons can be produced at LEP via the process $\epem \to
\gamma G$ and emitted in the extra dimensions carrying away energy.
This process
manifests itself through an enhancement of the single photon cross
section and a modification of the expected energy and polar angle
distributions of the
detected  photons~\cite{giudice,mirabelli}. The differential cross 
section of this process, as a function of the 
fraction $x_\gamma$ of the beam energy taken by the photon and its
polar emission 
angle $\theta$ relative to the beam, is given by~\cite{giudice}:
\begin{equation}
\frac{d^2\sigma(\epem \ra \gamma G)}{dx_\gamma d\cos \theta}
=\frac{\alpha}{64}~S_{\delta -1}
\left( \frac{\sqrt{s}}{M_D}\right)^{\delta +2}
\frac{1}{s}f(x_\gamma, \cos\theta, \delta),
\label{formula:ggravity}
\end{equation}

\noindent
where $\alpha$ is the QED coupling, chosen as  
$\alpha(\sqrt{s})$, the function $f(x,y,k)$ is given by:
\begin{equation}
f(x,y,k) = \frac{2(1-x)^{\frac{k}{2} -1}}{x(1-y^2)}
\left[ (2-x)^2(1-x+x^2)-3y^2x^2(1-x)-y^4x^4 \right],
\end{equation}

\noindent
and for $\delta=2n$ and $n$ integer,
$S_{\delta -1}=2\pi^n/(n-1)!$, while for $\delta=2n+1$, 
$S_{\delta -1}=2\pi^n/\Pi_{k=0}^{n-1} (k+\frac{1}{2})$.

Single photon events are selected at $183\GeV$
for an integrated
luminosity of $55.3\,\mathrm{pb}^{-1}$~\cite{l3photon}.
They
are characterised by missing energy and at least one detected photon of energy
greater than $4\GeV$ and polar angle above $14^{\circ}$. A minimal
transverse momentum of $5\GeV$ is required for  the photons
whose energy is greater than $5\GeV$, this cut being reduced to 
$2.7\GeV$ otherwise.  
The efficiency for a LSG signal is derived on a $x_\gamma-\cos{\theta}$  
grid from the   KORALZ~\cite{KORALZ}
$\epem\ra\nu\bar{\nu}\gamma(\gamma)$ Monte Carlo. Table~1 reports the
efficiencies as a function of $\delta$ for a signal  within the
described experimental energy and angular limits.  
The $x_\gamma$ spectrum expected from $\gamma G$ production and the SM
is then compared with data to derive upper limits at 
 95\% confidence level
(CL) on the cross section of this process
($\sigma_{\gamma G}^\mathrm{lim}$) and subsequently lower limits on
$M_D$ following  a Bayesian
approach. These limits are listed in Table~1  as a function of $\delta$.
Figure~1 shows the $x_\gamma$ spectra for data and SM Monte Carlo  together
with the modification expected from LSG with
$M_D=400\GeV$ and $\delta=6$.

\begin{table}[h]
  \begin{center}
    \begin{tabular}{|r|ccccccccc|}
      \hline
      $\delta$ & 2 & 3 & 4 & 5 & 6 & 7 & 8 & 9 & 10 \\ 
      \hline
      $\varepsilon$ (\%) & 43.4 & 41.4 & 39.7 & 38.3 & 37.1 & 36.1 & 35.2 &
      34.5 & 33.8 \\
      $\sigma_{\gamma G}^\mathrm{lim}$ (pb) & 0.79 & 0.76 &
      0.73 & 0.72 & 0.70 & 0.69 & 0.68 & 0.67 & 0.66 \\ 
      $M_D$ (GeV) & 945 & 769 & 644 & 555 & 489 & 439 & 399 & 367 & 340 \\ 
      \hline
    \end{tabular}
    \icaption[cuts]{Detection efficiencies for the real graviton
    plus photon signal, upper limit at 95\,\% CL on its cross
    section and lower limit on the scale $M_D$ as a function of 
    the number of extra dimensions.}
  \end{center}
\end{table}

%
%
\section{Virtual graviton effects}

 The production of $\gamma\gamma$, $\Zo\Zo$, $\Wp\Wm$,
$\mu^+\mu^-$ $\tau^+\tau^-$, $\qqbar$ and  $\epem$
  via a virtual graviton and its
interference with the SM description of those processes
can be analysed in term of the LSG cutoff energy,
$M_S$\footnote{The $M_S$ used here corresponds to the
  one of~\cite{hewett}, and is related to the $\Lambda_T$ cutoff
  of~\cite{giudice} by $\Lambda_T = (\pi/2 )^{1/4} M_S$.}, expected to be of the order of
$M_D$~\cite{giudice}. In the
following  radiative corrections to SM and LSG
processes are assumed to factorize.

%
%
\subsection{The \boldmath{$\epem\ra\gamma\gamma$} process}

The differential cross section for photon pair production in
$\epem$ collisions is modified by $s-$channel graviton exchange
as~\cite{giudice,agashe}:
\begin{equation}
 \frac{d \sigma(\epem\ra\gamma\gamma)}{d \cos \theta}
= \frac{\pi}{s} \left[ \alpha F_1 
\left (   - \sin ^2 \frac{\theta}{2}\right)  - \lambda
\frac{4 s^2}{\pi M_S^4} F_2 \left (  - \sin ^2 \frac{\theta}{2}\right)
\right] ^2,
\label{formula:gggravity}
\end{equation}
\noindent
where:
\begin{equation}
F_1 (x)  =  \sqrt{\frac{1 + 2 x + 2 x^2}{-x (1 +x)}},\,\,\,\mathrm{and}\,\,\,
F_2 (x)  =  \sqrt{\frac{-x (1 +x) ( 1 + 2 x + 2 x^2)}{16}},
\end{equation}
and the angle $\theta$ is the photon emission angle with respect to
the beam axis. The LSG contribution is weighted by a further factor
$\lambda$~\cite{hewett}, which 
incorporates other possible model dependences. It is chosen as
$\pm 1$ in the following to allow for different signs in the
interference.

The  process $\epem\ra\gamma\gamma$ is investigated 
at $\sqrt{s} = 130\GeV - 140\GeV$~\cite{l3gg140}, $\sqrt{s} = 161\GeV -
172\GeV$~\cite{l3gg161} and $\sqrt{s} = 183\GeV$~\cite{l3gg183}. 
Figure~2a shows the distribution of the polar
angle of the events selected in this total sample. 
A fit to this distribution for each energy point with 
the relation~(\ref{formula:gggravity}) yields the likelihood curve  
presented in Figure~3  as a function of $\lambda/M_S^4$. This result  is
compatible with SM predictions and allows the
extraction of a 95\,\% CL limit on $M_S$ of $630\GeV$ for $\lambda
= +1$ and $670\GeV$ for $\lambda = -1$, by integrating the likelihood
over the physical region of positive $M_S^4$. These limits are unaffected
by the estimated 1\% systematic uncertainty on the measured
differential cross section. In this and in the
following fits, the background dependence on LSG effects is negligible.

%
%
\subsection{The \boldmath{$\epem\ra\Zo\Zo$} process}

The contribution of virtual graviton exchange to the pair production
of Z bosons is given in Reference~\cite{agashe}. The total cross section
of this process at $\sqrt{s} =183\GeV$ is expected to change by 43\%
for $M_S$ equal to $500\GeV$.

The same distributions used to derive limits~\cite{l3zz} on the triple
neutral gauge boson anomalous couplings 
$f^{\mathrm{Z},\gamma}_{4,5}$ are used to investigate possible LSG
contributions to this process. The EXCALIBUR Monte
Carlo program~\cite{EXCALIBUR} is modified  to include  the  matrix
element for graviton exchange~\cite{agashe}. 
Monte Carlo events  are then reweighted as a function of 
$\lambda/M_S^4$ and compared to data  yielding the likelihood curve
presented in Figure~3. No 
deviations from the SM are observed, giving 95\,\% CL limits on
$M_S$ of $470\GeV$ and $460\GeV$ for  $\lambda = +1$ and $\lambda =
-1$ respectively.  Systematic effects are found to be negligible.

%
%
\subsection{The \boldmath{$\epem\ra\Wp\Wm$} process}

The LSG contributions to W pair
production  modify the differential cross section~\cite{agashe}.
The polar angle of the emitted $\Wm$ boson
for semileptonic and hadronic decays of the W pair is
studied~\cite{l3ww}.
The Monte Carlo events  are
reweighted to model  LSG effects using  a modified
version of the EXCALIBUR Monte
Carlo program which includes the
virtual graviton exchange matrix element~\cite{agashe} as a
function of  $\lambda/M_S^4$. Only the effects for
the double--resonant processes are taken into account. A 5\%
correction is applied for other  diagrams for the semileptonic
electron final states.
Figure~2b presents the $\cos{\theta}$ distributions for
data, background and reconstructed
$\epem\ra\Wp\Wm\ra\ell\nu_{\ell}\rm{q\bar{ q}'}$ events as expected
from the SM Monte Carlo alone and with the addition of LSG effects.

Separate fits are  performed on the 
reconstructed $\cos{\theta}$ distributions for each of the three
semileptonic classes of events and the hadronic one. 
The combined likelihood curve is plotted in
Figure~3. This curve is compatible with 
the hypothesis of no contribution from extra dimensions to the
$\epem\ra\Wp\Wm$ process and 95\% CL limits on $M_S$ can thus be extracted  
as $650\GeV$ if  $\lambda = +1$ and $520\GeV$ if   $\lambda =
-1$. This procedure takes into account 7\% experimental and 2\%
theoretical systematic uncertainties on the total cross section.

%
%
\subsection{The \boldmath{$\epem\ra f \bar{f}$} process}

The effect of the  $s$--channel graviton exchange in fermion pair
production  other than Bhabha scattering is described
in References~\cite{hewett,giudice}. In the massless fermion limit the formula
reads~\cite{hewett}:
\begin{eqnarray}
 \frac{d \sigma(\epem\ra f \bar{f})}{d \cos \theta}
& = & \frac{d \sigma^{SM}(\epem\ra f \bar{f})}{d \cos \theta}
+
 \frac{N_c \alpha s \lambda}{4 M_S^4} \nonumber \\
&\times &\left\{ 
2 Q_e Q_f \cos^3{\theta} +
\frac{s(s-M_Z^2)}{(s-M_Z^2)^2+M_Z^2\Gamma_Z^2}
\left[(2 v_e v_f \cos^3{\theta} - a_e a_f (1-3 \cos^2{\theta})
\right]
\right\} \nonumber \\ 
&+ &\frac{N_c s^3 \lambda^2}{32 \pi M_S^8}
(1 - 3 \cos^2{\theta} + 4 \cos^4{\theta}),
\label{formula:ffbar}
\end{eqnarray}
where $\theta$ is the fermion production angle with respect to the
incoming electron, $Q_i$, $a_i$ and $v_i$ are the charge, axial and
vector couplings of the fermion $i$,  $\Gamma_Z$ and $M_Z$ stand for the Z
width and mass and $N_c$ is the number of colours of the fermion $f$. The first
term 
is the SM cross section which 
is calculated with  ZFITTER~\cite{ZFITTER}.

Full energy muon and tau pairs are selected at
$\sqrt{s}=183\GeV$~\cite{ffbar183}.  
The polar angle distribution of
the identified negative charged lepton is compared in Figures~2c
and~2d with 
Monte Carlo events simulated with KORALZ.
The LSG  effects are included by 
reweighting. The expected effects  for $\lambda=\pm 1$ and $M_S
= 450\GeV$ are also presented in  Figures~2c and~2d.

From  the muon distribution 95\%
CL lower limits
on $M_S$ are derived as $550\GeV$ and $490\GeV$ for  $\lambda=+1$ and
$\lambda=-1$, respectively. The investigation of tau pairs yields
the corresponding limits of $510\GeV$ and $460\GeV$ at 95\%
CL. The experimental and theoretical systematic
uncertainties on the total cross sections amount
to 2.4\% for muons and 3.5\% for taus. They are included in the
derivation of the limit.
 
For the $\qqbar$ final
states  only the integrated cross
section is investigated, where the higher sensitivity
interference term proportional to  $1/M_S^4$ vanishes, leaving the
$1/M_S^8$ factor of the pure 
graviton exchange. The samples collected for full energy
 $\qqbar$ final states at $\sqrt{s}$ of 
$130\GeV-136\GeV$~\cite{ffbar130136}, $161\GeV-172\GeV$~\cite{ffbar161172} and
$183\GeV$~\cite{ffbar183} are studied. The large statistics, the
enhancement due to the colour factor 
and the sum over five flavours, allow  to set lower limits
on $M_S$ as high as $490\GeV$ at  95\%
CL, independent of the sign of $\lambda$. A 
systematic uncertainty of 1.4\% is included in the limit calculation.

For Bhabha scattering the differential cross
section in presence of LSG is~\cite{giudice,rizzo}:
\begin{eqnarray}
 \frac{d \sigma(\epem\ra \epem)}{d \cos \theta}
& = & \frac{d \sigma^{SM}(\epem\ra\epem)}{d \cos \theta}
-
 \frac{\alpha \lambda}{2 s M_S^4} \nonumber \\
&\times &\left\{ 
{\cal{F}}_1(s,t)+
\frac{
v_e^2{\cal{F}}_2(s,t)+
a_e^2{\cal{F}}_3(s,t)
}{s-M_Z^2}
+
\frac{
v_e^2{\cal{F}}_2(t,s)+
a_e^2{\cal{F}}_3(t,s)
}{t-M_Z^2}
\right\}
\nonumber \\ 
&+ &\frac{\lambda^2}{16 \pi s M_S^8}
{\cal{F}}_4(s,t)
,
\label{formula:bhabha}
\end{eqnarray}
where the functions ${\cal{F}}_i$ of $s$ and $t$ are written as:
\begin{eqnarray}
{\cal{F}}_1(s,t)&=& 9\left({s^3 /  {t}}+{t^3 / {s}}\right)+23(s^2+t^2)+30st
\,, \nonumber \\
{\cal{F}}_2(s,t)&=& 5s^3+10s^2t+18st^2+9t^3\,, \nonumber \\
{\cal{F}}_3(s,t)&=& 5s^3+15s^2t+12st^2+t^3\,, \nonumber \\
{\cal{F}}_4(s,t)&=& 41(s^4+t^4)+124st(s^2+t^2)+148s^2t^2\,.
\end{eqnarray}
The differential cross section for Bhabha scattering at $183\GeV$ for
full energy events between 44$^\circ$ and 136$^\circ$ of polar
angle~\cite{ffbar183} is compared with equation
(\ref{formula:bhabha}). The first term of this equation represents the SM
predictions, calculated using the  BHWIDE~\cite{bhwide} Monte Carlo  
and normalised to the TOPAZ0~\cite{topaz0} semi--analytical
calculation.  This yields limits on $M_S$ of $810\GeV$
and $720\GeV$ at 95\% CL
for  $\lambda=+1$ and $\lambda=-1$, respectively. Experimental and
theoretical systematic uncertainties are included in the fit and amount in
total to 3.7\%.
Figure~4 shows the measured and expected SM differential
cross sections and their differences, together with the  LSG
predictions.

A simultaneous fit to all the fermion pair channels  improves the $\lambda=+1$
limit to $820\GeV$. The likelihood
curve of this combined fit is  displayed in Figure~3 as a function of $\lambda/M_S^4$.

%
%
\subsection{Combined Results}

Assuming that no  higher order operators give sizeable
contributions to the Equations~(\ref{formula:gggravity}), 
(\ref{formula:ffbar}) and (\ref{formula:bhabha}), and to the LSG ZZ and
$\Wp\Wm$ matrix elements and that the meaning of the
cutoff parameter is the same for all the investigated processes, it is possible to combine
the likelihood curves obtained for all pair production
processes into a single one, also displayed in Figure~3.

No indication of the contribution of virtual
graviton exchange is found and lower limits at 95\% CL on the value
of the scale 
$M_S$ are derived as $860\GeV$ for $\lambda = +1$ and
$740\GeV$ for $\lambda = -1$.  The  individual and combined 
limits are summarised in Table~2.

\begin{table}[ht]
  \begin{center}
    \begin{tabular}{|c|c|c|}
       \hline
       Process                & $M_S$\,(GeV)& $M_S$\,(GeV)\\
                              & $\lambda = +1$ & $\lambda = -1$ \\
       \hline
       $\epem\ra\gamma\gamma$   &  630 & 670\\
       $\epem\ra\Zo\Zo$         &  470 & 460\\
       $\epem\ra\Wp\Wm$         &  650 & 520\\
       \hline                               
       Bosons Combined          &  700 & 670\\
       \hline                               
       $\epem\ra\qqbar    $     &  490 & 490\\
       $\epem\ra\epem     $     &  810 & 720\\
       $\epem\ra\mu^+\mu^-$     &  550 & 490\\
       $\epem\ra\tau^+\tau^-$   &  510 & 460\\
       \hline                               
       Fermions Combined        &  820 & 720\\
       \hline                               
       Bosons + Fermions        &  860 & 740\\
       \hline
    \end{tabular}
    \icaption[cuts]{Lower limits at 95\,\% CL on the cutoff
    $M_S$ for different processes and values of
    $\lambda$. 
    Combined results are also given.}
  \end{center}
\end{table}

%
%
\section*{Acknowledgements}

We  thank  Roberto Pittau for his kind help 
in including Low Scale Gravity effects
in  EXCALIBUR and  Gian Giudice and Enrique Alvarez for clarifying discussions
on extra dimensions  theory issues.
We wish to express our gratitude to the CERN accelerator divisions for the
excellent performance and the continuous and successful upgrade of the
LEP machine.  
We acknowledge the contributions of the engineers  and technicians who
have participated in the construction and maintenance of this experiment.

%
%

\bibliographystyle{l3stylem}

%
%
\newpage
\typeout{   }     
\typeout{Using author list for paper 180 -?}
\typeout{$Modified: Wed Aug 25 10:07:39 1999 by clare $}
\typeout{!!!!  This should only be used with document option a4p!!!!}
\typeout{   }
%
%
%
%
%
%

\newcount\tutecount  \tutecount=0
\def\tutenum#1{\global\advance\tutecount by 1 \xdef#1{\the\tutecount}}
\def\tute#1{$^{#1}$}
\tutenum\aachen            
\tutenum\nikhef            
\tutenum\mich              
\tutenum\lapp              
\tutenum\basel             
\tutenum\lsu               
\tutenum\beijing           
\tutenum\berlin            
\tutenum\bologna           
\tutenum\tata              
\tutenum\ne                
\tutenum\bucharest         
\tutenum\budapest          
\tutenum\mit               
\tutenum\debrecen          
\tutenum\florence          
\tutenum\cern              
\tutenum\wl                
\tutenum\geneva            
\tutenum\hefei             
\tutenum\seft              
\tutenum\lausanne          
\tutenum\lecce             
\tutenum\lyon              
\tutenum\madrid            
\tutenum\milan             
\tutenum\moscow            
\tutenum\naples            
\tutenum\cyprus            
\tutenum\nymegen           
\tutenum\caltech           
\tutenum\perugia           
\tutenum\cmu               
\tutenum\prince            
\tutenum\rome              
\tutenum\peters            
\tutenum\salerno           
\tutenum\ucsd              
\tutenum\santiago          
\tutenum\sofia             
\tutenum\korea             
\tutenum\alabama           
\tutenum\utrecht           
\tutenum\purdue            
\tutenum\psinst            
\tutenum\zeuthen           
\tutenum\eth               
\tutenum\hamburg           
\tutenum\taiwan            
\tutenum\tsinghua          
{
\parskip=0pt
\noindent
{\bf The L3 Collaboration:}
\ifx\selectfont\undefined
 \baselineskip=10.8pt
 \baselineskip\baselinestretch\baselineskip
 \normalbaselineskip\baselineskip
 \ixpt
\else
 \fontsize{9}{10.8pt}\selectfont
\fi
\medskip
\tolerance=10000
\hbadness=5000
\raggedright
\hsize=162truemm\hoffset=0mm
\def\r{\rlap,}
\noindent

M.Acciarri\r\tute\milan\
P.Achard\r\tute\geneva\ 
O.Adriani\r\tute{\florence}\ 
M.Aguilar-Benitez\r\tute\madrid\ 
J.Alcaraz\r\tute\madrid\ 
G.Alemanni\r\tute\lausanne\
J.Allaby\r\tute\cern\
A.Aloisio\r\tute\naples\ 
M.G.Alviggi\r\tute\naples\
G.Ambrosi\r\tute\geneva\
% ...
\newpage
\begin{list}{A}{\itemsep=0pt plus 0pt minus 0pt\parsep=0pt plus 0pt minus 0pt
                \topsep=0pt plus 0pt minus 0pt}
\item[\aachen]
 I. Physikalisches Institut, RWTH, D-52056 Aachen, FRG$^{\S}$\\
 III. Physikalisches Institut, RWTH, D-52056 Aachen, FRG$^{\S}$
\item[\nikhef] National Institute for High Energy Physics, NIKHEF, 
     and University of Amsterdam, NL-1009 DB Amsterdam, The Netherlands
\item[\mich] University of Michigan, Ann Arbor, MI 48109, USA
\item[\lapp] Laboratoire d'Annecy-le-Vieux de Physique des Particules, 
     LAPP,IN2P3-CNRS, BP 110, F-74941 Annecy-le-Vieux CEDEX, France
\item[\basel] Institute of Physics, University of Basel, CH-4056 Basel,
     Switzerland
\item[\lsu] Louisiana State University, Baton Rouge, LA 70803, USA
\item[\beijing] Institute of High Energy Physics, IHEP, 
  100039 Beijing, China$^{\triangle}$ 
\item[\berlin] Humboldt University, D-10099 Berlin, FRG$^{\S}$
\item[\bologna] University of Bologna and INFN-Sezione di Bologna, 
     I-40126 Bologna, Italy
\item[\tata] Tata Institute of Fundamental Research, Bombay 400 005, India
\item[\ne] Northeastern University, Boston, MA 02115, USA
\item[\bucharest] Institute of Atomic Physics and University of Bucharest,
     R-76900 Bucharest, Romania
\item[\budapest] Central Research Institute for Physics of the 
     Hungarian Academy of Sciences, H-1525 Budapest 114, Hungary$^{\ddag}$
\item[\mit] Massachusetts Institute of Technology, Cambridge, MA 02139, USA
\item[\debrecen] Lajos Kossuth University-ATOMKI, H-4010 Debrecen, Hungary$^\P$
\item[\florence] INFN Sezione di Firenze and University of Florence, 
     I-50125 Florence, Italy
\item[\cern] European Laboratory for Particle Physics, CERN, 
     CH-1211 Geneva 23, Switzerland
\item[\wl] World Laboratory, FBLJA  Project, CH-1211 Geneva 23, Switzerland
\item[\geneva] University of Geneva, CH-1211 Geneva 4, Switzerland
\item[\hefei] Chinese University of Science and Technology, USTC,
      Hefei, Anhui 230 029, China$^{\triangle}$
\item[\seft] SEFT, Research Institute for High Energy Physics, P.O. Box 9,
      SF-00014 Helsinki, Finland
\item[\lausanne] University of Lausanne, CH-1015 Lausanne, Switzerland
\item[\lecce] INFN-Sezione di Lecce and Universit\'a Degli Studi di Lecce,
     I-73100 Lecce, Italy
\item[\lyon] Institut de Physique Nucl\'eaire de Lyon, 
     IN2P3-CNRS,Universit\'e Claude Bernard, 
     F-69622 Villeurbanne, France
\item[\madrid] Centro de Investigaciones Energ{\'e}ticas, 
     Medioambientales y Tecnolog{\'\i}cas, CIEMAT, E-28040 Madrid,
     Spain${\flat}$ 
\item[\milan] INFN-Sezione di Milano, I-20133 Milan, Italy
\item[\moscow] Institute of Theoretical and Experimental Physics, ITEP, 
     Moscow, Russia
\item[\naples] INFN-Sezione di Napoli and University of Naples, 
     I-80125 Naples, Italy
\item[\cyprus] Department of Natural Sciences, University of Cyprus,
     Nicosia, Cyprus
\item[\nymegen] University of Nijmegen and NIKHEF, 
     NL-6525 ED Nijmegen, The Netherlands
\item[\caltech] California Institute of Technology, Pasadena, CA 91125, USA
\item[\perugia] INFN-Sezione di Perugia and Universit\'a Degli 
     Studi di Perugia, I-06100 Perugia, Italy   
\item[\cmu] Carnegie Mellon University, Pittsburgh, PA 15213, USA
\item[\prince] Princeton University, Princeton, NJ 08544, USA
\item[\rome] INFN-Sezione di Roma and University of Rome, ``La Sapienza",
     I-00185 Rome, Italy
\item[\peters] Nuclear Physics Institute, St. Petersburg, Russia
\item[\salerno] University and INFN, Salerno, I-84100 Salerno, Italy
\item[\ucsd] University of California, San Diego, CA 92093, USA
\item[\santiago] Dept. de Fisica de Particulas Elementales, Univ. de Santiago,
     E-15706 Santiago de Compostela, Spain
\item[\sofia] Bulgarian Academy of Sciences, Central Lab.~of 
     Mechatronics and Instrumentation, BU-1113 Sofia, Bulgaria
\item[\korea] Center for High Energy Physics, Adv.~Inst.~of Sciences
     and Technology, 305-701 Taejon,~Republic~of~{Korea}
\item[\alabama] University of Alabama, Tuscaloosa, AL 35486, USA
\item[\utrecht] Utrecht University and NIKHEF, NL-3584 CB Utrecht, 
     The Netherlands
\item[\purdue] Purdue University, West Lafayette, IN 47907, USA
\item[\psinst] Paul Scherrer Institut, PSI, CH-5232 Villigen, Switzerland
\item[\zeuthen] DESY, D-15738 Zeuthen, 
     FRG
\item[\eth] Eidgen\"ossische Technische Hochschule, ETH Z\"urich,
     CH-8093 Z\"urich, Switzerland
\item[\hamburg] University of Hamburg, D-22761 Hamburg, FRG
\item[\taiwan] National Central University, Chung-Li, Taiwan, China
\item[\tsinghua] Department of Physics, National Tsing Hua University,
      Taiwan, China
\item[\S]  Supported by the German Bundesministerium 
        f\"ur Bildung, Wissenschaft, Forschung und Technologie
\item[\ddag] Supported by the Hungarian OTKA fund under contract
numbers T019181, F023259 and T024011.
\item[\P] Also supported by the Hungarian OTKA fund under contract
  numbers T22238 and T026178.
\item[$\flat$] Supported also by the Comisi\'on Interministerial de Ciencia y 
        Tecnolog{\'\i}a.
\item[$\sharp$] Also supported by CONICET and Universidad Nacional de La Plata,
        CC 67, 1900 La Plata, Argentina.
\item[$\diamondsuit$] Also supported by Panjab University, Chandigarh-160014, 
        India.
\item[$\triangle$] Supported by the National Natural Science
  Foundation of China.
\item[\dag] Deceased.
\end{list}
}
\vfill





\newpage

%
%

\begin{figure}[p]
  \begin{center}
      \mbox{\includegraphics[width=\figwidth]{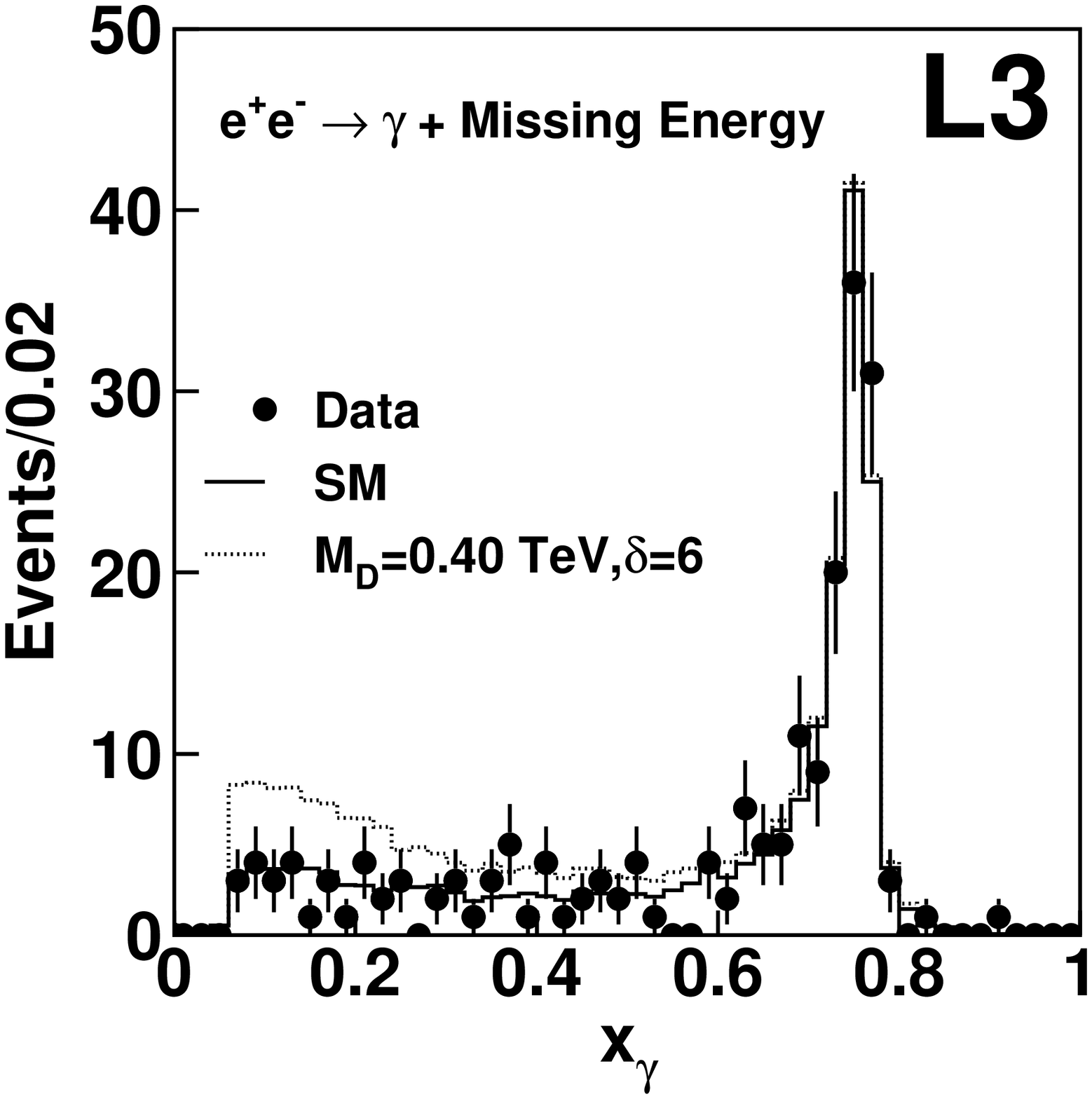}} 
    \icaption{Distribution of ratio of the photon energy to the beam
    energy for single photon events in data at $\sqrt{s}=183\GeV$
    together with SM expectations. The
    effect of 
    real graviton production with six extra space dimensions and
    $M_D=0.4\TeV$ is also shown.} 
    \label{fig:fig1}
  \end{center}
\end{figure}

\begin{figure}[p]
  \begin{center}
    \begin{tabular}{cc}
      \mbox{\includegraphics[width=.5\figwidth]{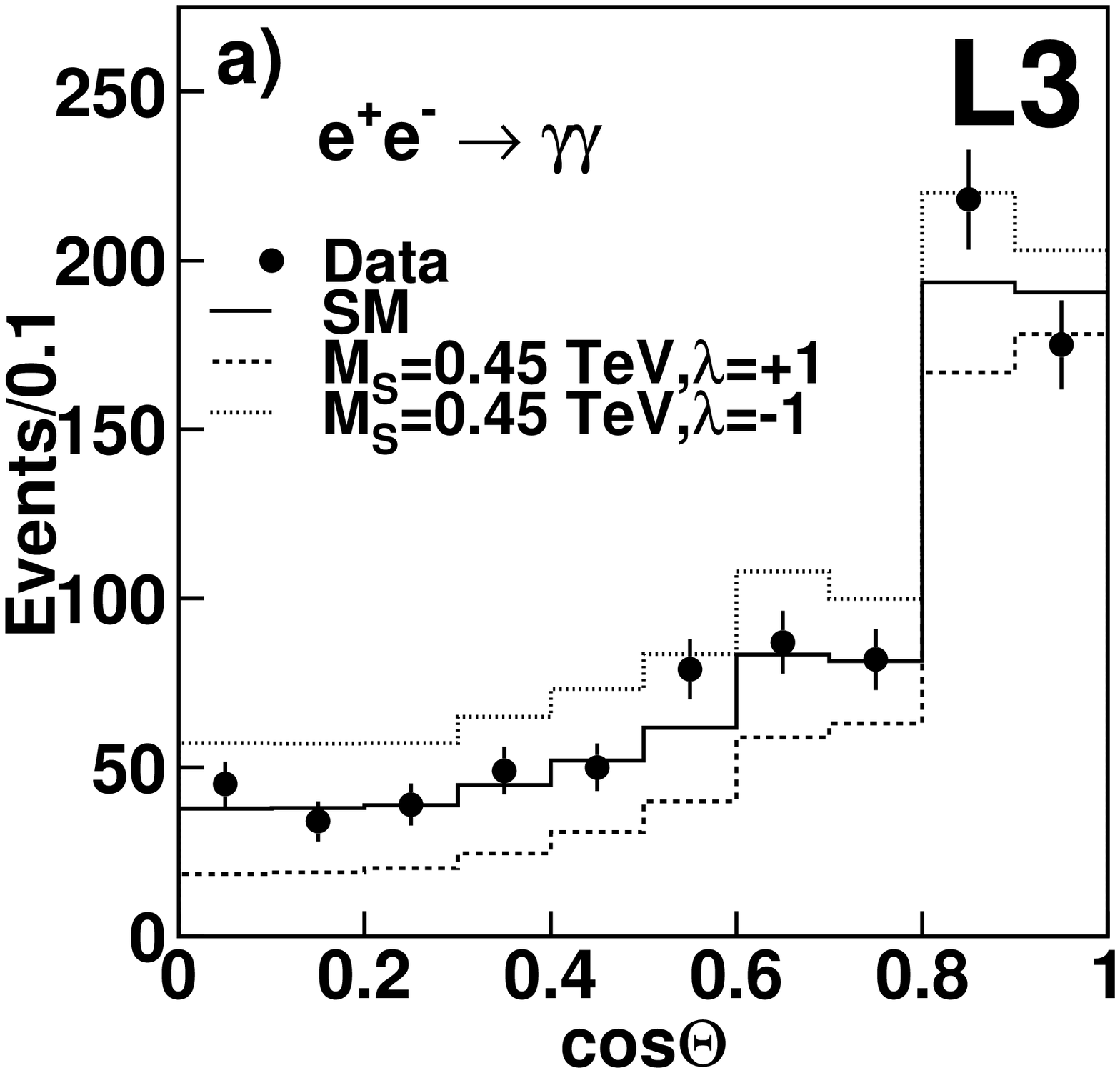}} &
      \mbox{\includegraphics[width=.5\figwidth]{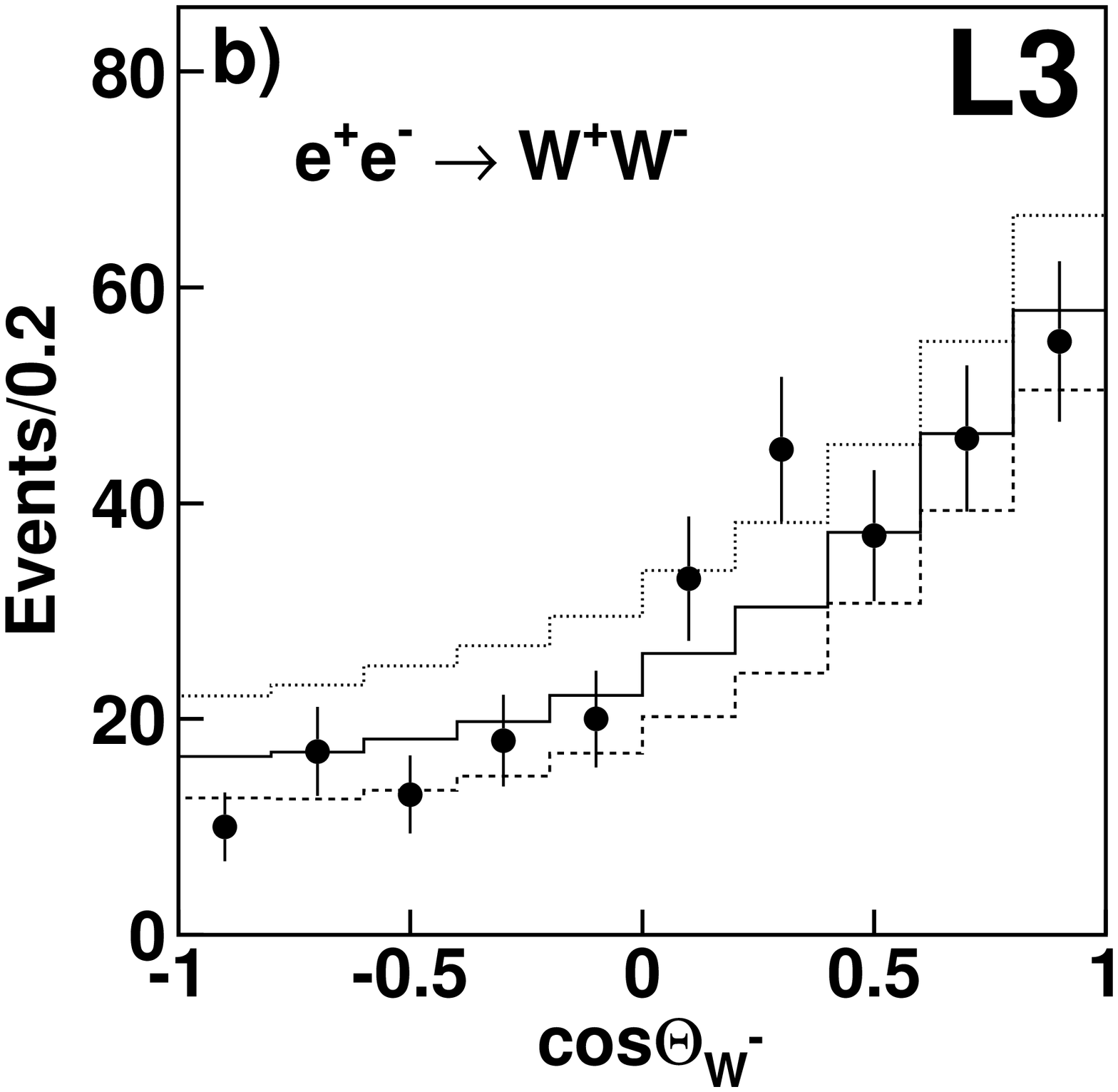}} \\
      \mbox{\includegraphics[width=.5\figwidth]{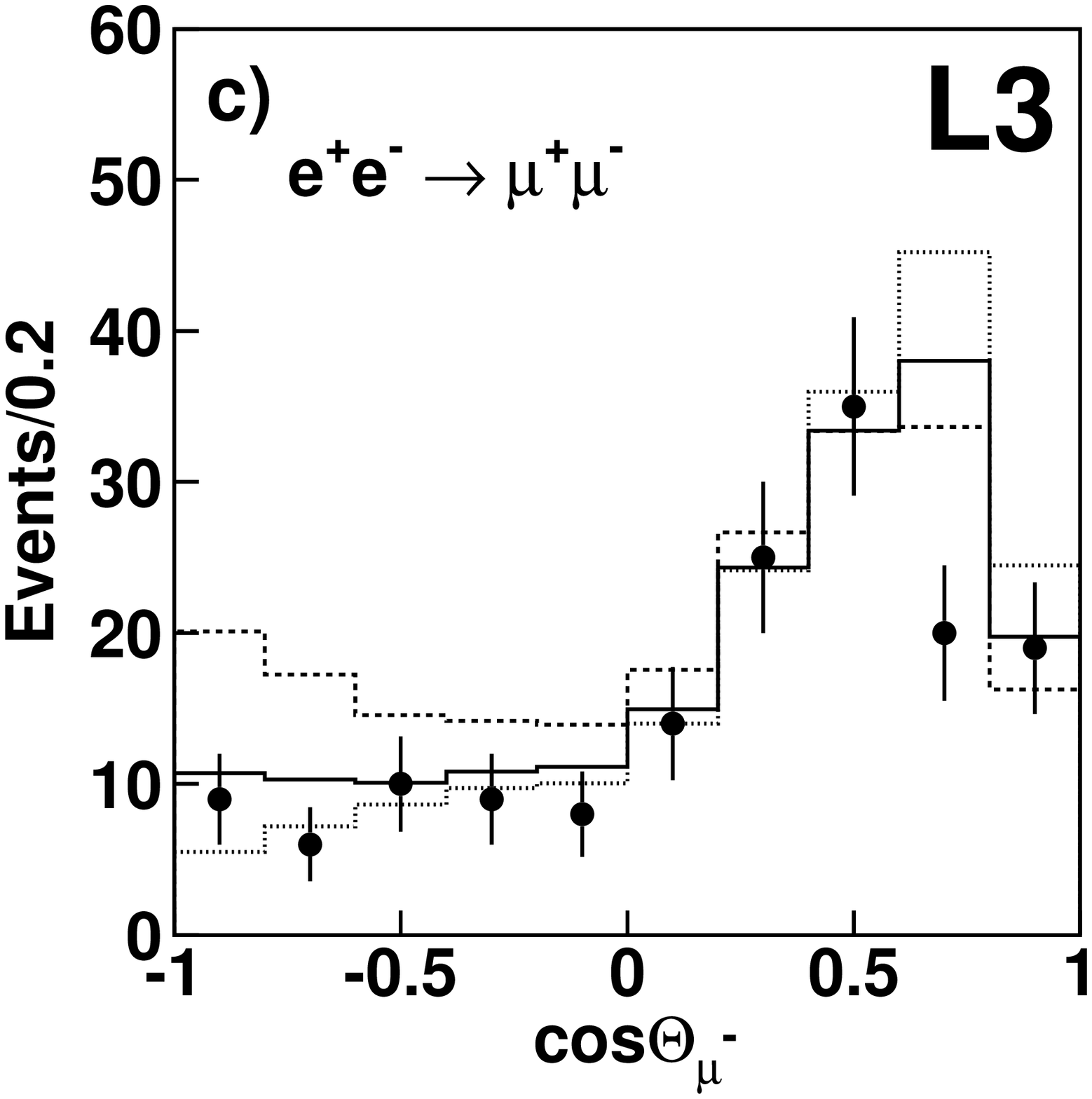}} &
      \mbox{\includegraphics[width=.5\figwidth]{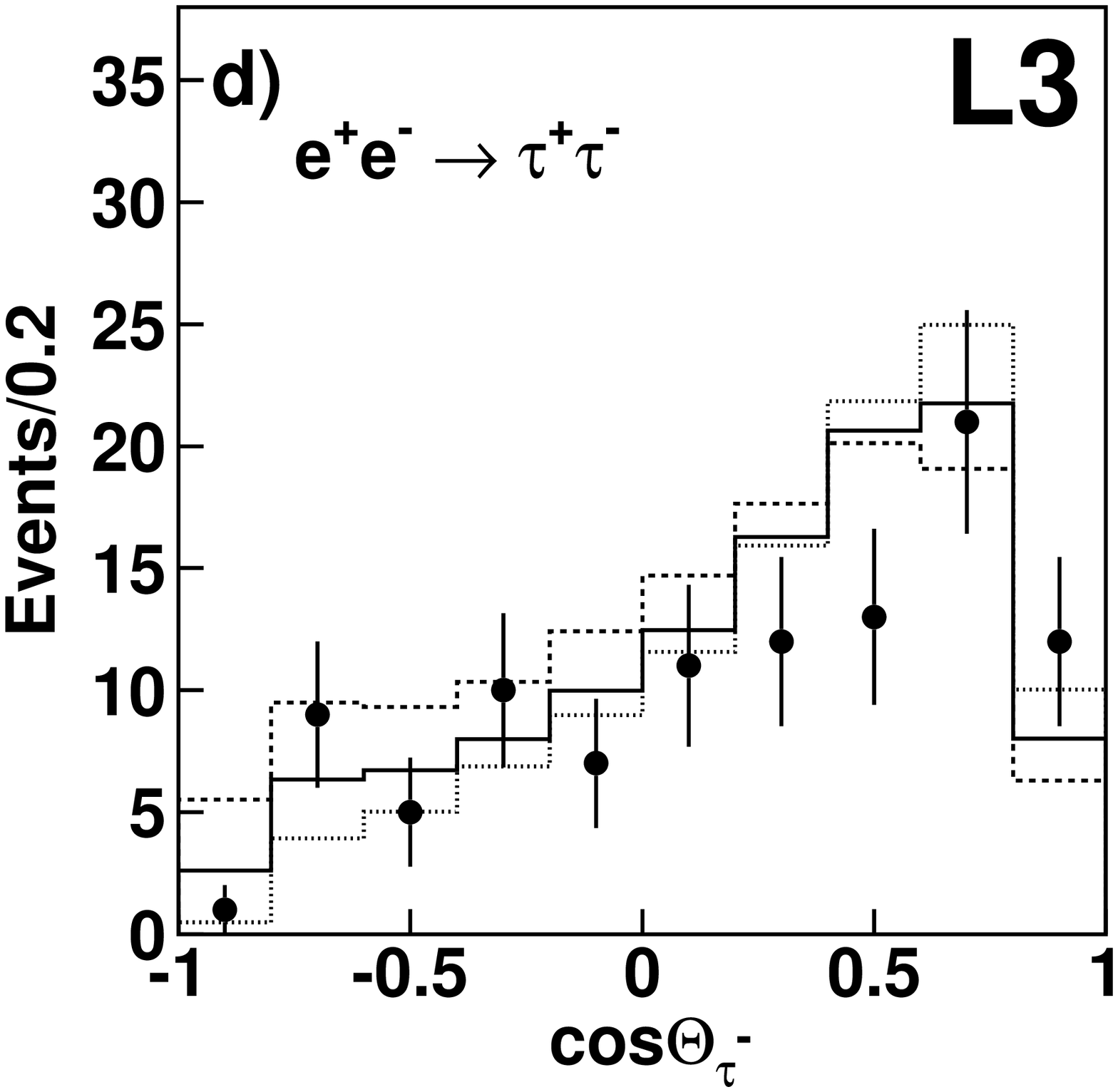}} \\
    \end{tabular}
    \icaption{Distributions of the polar angle for: (a) 
    $\epem\ra\gamma\gamma$ events  selected at $\sqrt{s} =
    130-183\GeV$, (b)  semileptonic
    $\epem\ra\Wp\Wm$ events, (c)  $\epem\ra\mu^+\mu^-$ 
    and (d)  $\epem\ra\tau^+\tau^-$ processes. The last three plots refer to
    the $\sqrt{s} 
    =183\GeV$ data sample only. Data, SM expectations
    and LSG effects for $M_S=0.45\TeV$ are shown.}
    \label{fig:fig2}
  \end{center}
\end{figure}

\begin{figure}[p]
  \begin{center}
      \mbox{\includegraphics[width=\figwidth]{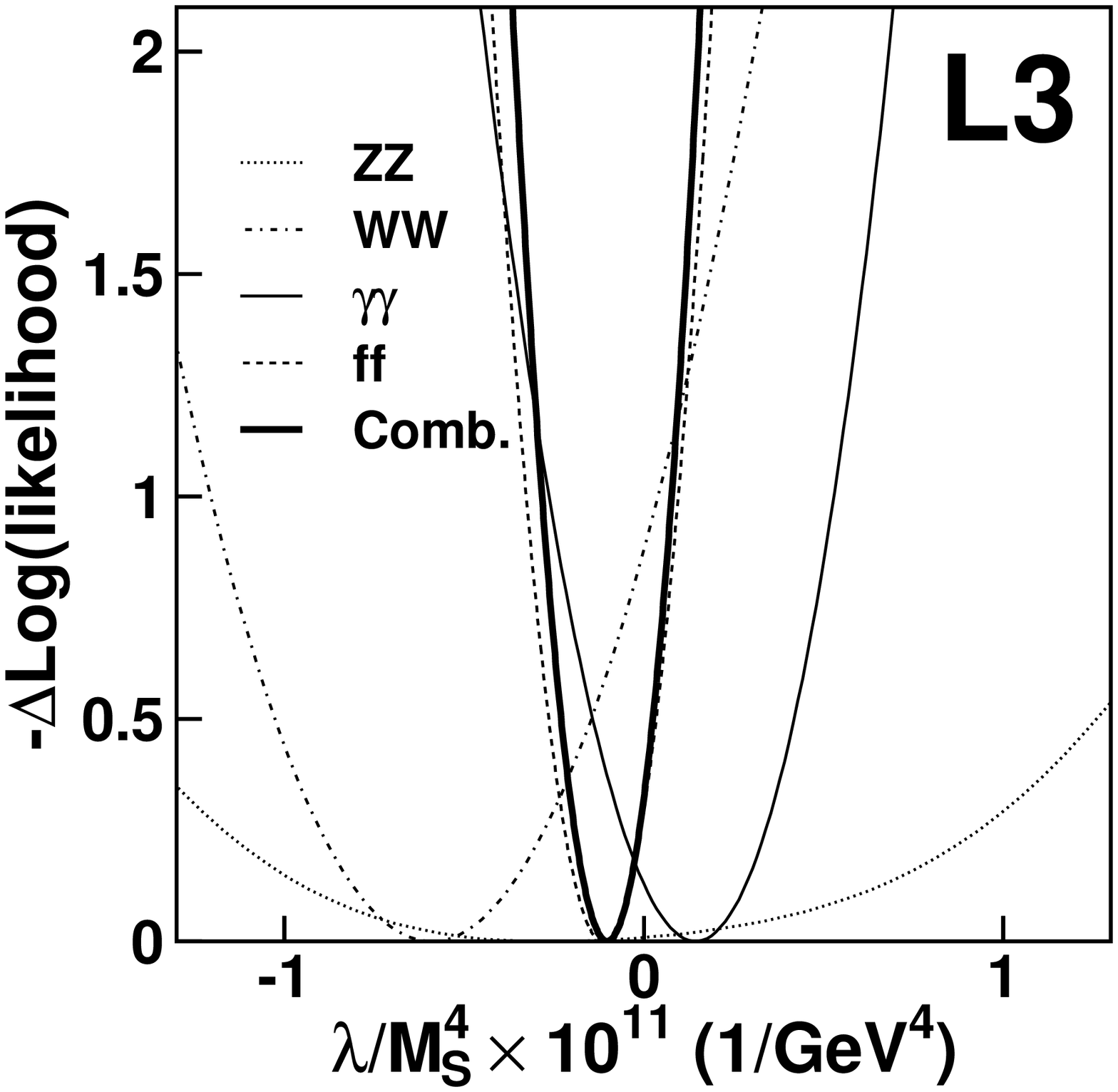}}
    \icaption{Likelihood curves for ${\lambda \over  M_S^4}$ for
      all the processes described in the text and their combination.}
    \label{fig:fig3}
  \end{center}
\end{figure}

\begin{figure}[p]
  \begin{center}
      \mbox{\includegraphics[width=\figwidth]{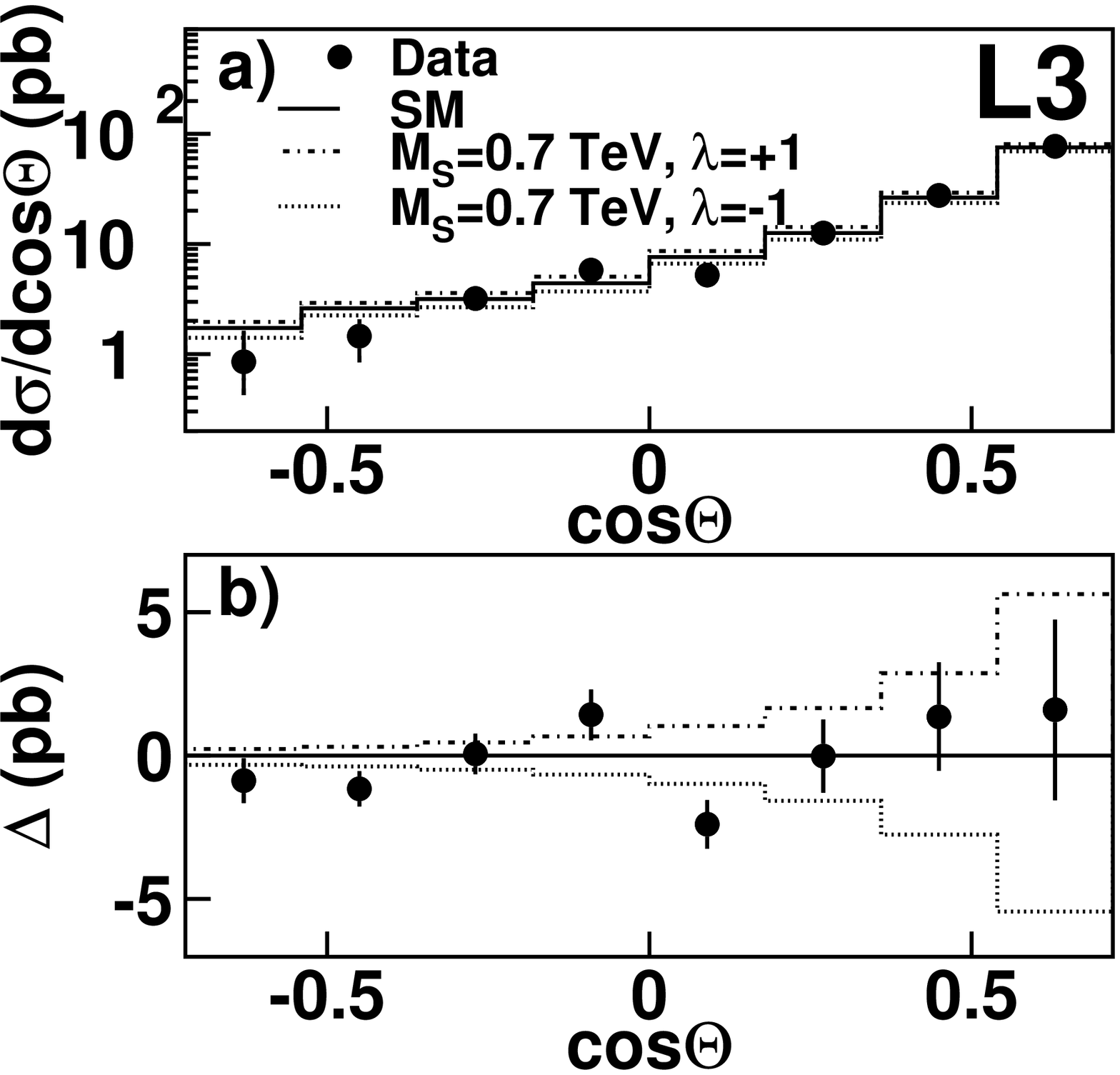}}
    \icaption{a) Differential cross section for Bhabha scattering in
    data and the SM expectations. LSG
    modifications for $M_S = 0.7\TeV$ are also displayed for
    $\lambda=\pm 1$. b) Differences $\Delta$ of the above with SM prediction.}
    \label{fig:fig4}
  \end{center}
\end{figure}

\end{document}